\documentclass[aps,prb,twocolumn,preprintnumbers,amsmath,amssymb,superscriptaddress]{revtex4}

\usepackage{graphicx}
\usepackage{dcolumn}
\usepackage{bm}
\newcommand{\nc}{\newcommand}
\nc{\be}{\begin{equation}}
\nc{\ee}{\end{equation}}
\nc{\bea}{\begin{eqnarray}}
\nc{\eea}{\end{eqnarray}}
\nc{\bean}{\begin{eqnarray*}}
\nc{\eean}{\end{eqnarray*}}
\nc{\mb}{\mbox}
\nc{\rnc}{\renewcommand}
\nc{\vk}{\mb{\bf k}}
\nc{\vp}{\mb{\bf p}}
\nc{\vn}{\mb{\bf n}}
\nc{\vq}{\mb{\bf q}}
\nc{\rr}{\mb{\bf r}}
\nc{\vz}{\hat {\mb{\bf z}}}
\nc{\vj}{\mb{\boldmath$j$}}
\nc{\vg}{\mb{\boldmath$g$}}
\nc{\x}{\mb{\boldmath$x$}}
\nc{\A}{\mb{\boldmath$A$}}
\nc{\va}{\mb{\boldmath$a$}}
\nc{\vs}{\mb{\boldmath$\sigma$}}
\nc{\vpi}{\mb{\boldmath$\pi$}}
\nc{\nab}{\nabla}
\nc{\X}{\sf x}

\begin{document}

\title{Non-Fermi Liquid behavior in Neutral Bilayer Graphene}

\author{Yafis Barlas}
\affiliation{National High Magnetic Field Laboratory and Department of Physics, Florida State
University, FL 32306, USA}
\author{Kun Yang}
\affiliation{National High Magnetic Field Laboratory and Department of Physics, Florida State
University, FL 32306, USA}

\begin{abstract}

We calculate the density-density response function and electron
self-energy for undoped bilayer graphene, within the Random Phase Approximation (RPA). We
show that the quasiparticle decay rate scales linearly with the quasiparticle energy, and
quasiparticle weight vanishes logarithmically in the low-energy limit, indicating non-Fermi
liquid behavior. This is a consequence of the absence of a Fermi surface for neutral bilayer
graphene and corresponding larger phase space available for scattering processes. Experimental 
consequences of our results as well as their differences from those of single-layer graphene are
discussed.

\end{abstract}

\maketitle


{\em Introduction} ---
The isolation and subsequent identification of graphene, an atomically thin electron system,
has
led to intense experimental and theoretical interest~\cite{reviews}. Recent experimental
progress~\cite{reviews} has also led to
techniques that enable isolation and study of systems with a small number of graphene layers, of
particular importance is AB-Bernal stacked bilayer graphene~\cite{bilayerexp}, a system which
shares some features both with graphene and two dimensional electron gas (2DEGs)~\cite{falko},
however at the same time different from both. Neglecting trigonal warping
balanced bilayer graphene can be identified as a zero-gap semiconductor with
quadratic dispersion; for undoped bilayer graphene the Fermi energy lies at the
neutral Fermi point, described as the point where the degenerate particle-hole bands meet.
Collectively these systems can be classified as chiral 2DEGs~\cite{hongki}. Electron-electron
interactions in chiral 2DEGs can lead to interesting quasiparticle properties, for example
quasiparticle velocity enhancement~\cite{guineaprbrapid} in graphene due to the presence of
unscreened Coulomb interactions. Most of the physics in this paper focuses on the difference
in graphene and bilayer graphene's chiral 2DEG.\newline

\begin{figure}[t]
\begin{center}
\hspace{-1cm}
\includegraphics[clip,width=3.40in,height=3.00in]{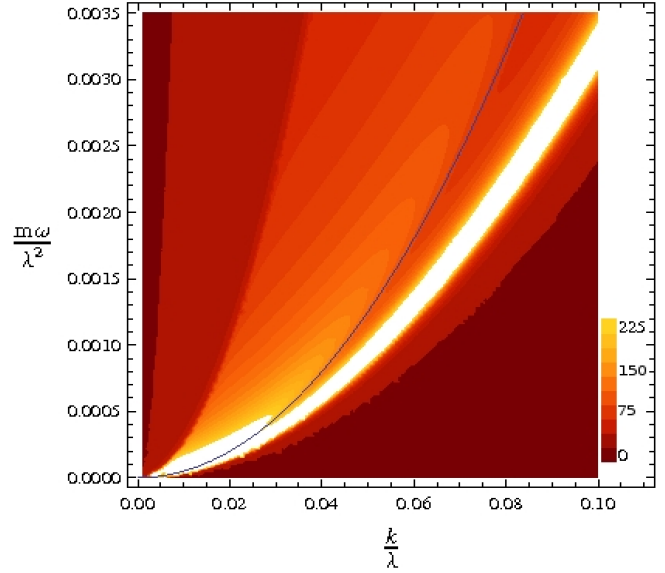}
\caption{(Color online) Intensity plot of electron spectral function at long-wavelengths and 
low-energies in units of $m/\lambda^{2}$.
The solid line corresponds to the noninteracting dispersion. $m$ is the electron effective mass and
and $\lambda= me^2/\varepsilon$ is proportional to $q_{TF}$ the Thomas-Fermi screening wave-vector.
The quasiparticle width (or scattering rate) is proportional to its energy and the quasiparticle
spectral weight vanishes logarithmically at low-energies. See text for details.}
\label{spectralfunction}
\end{center}
\end{figure}

\indent
In this paper we investigate the quasiparticle properties of undoped bilayer
graphene due to electron-electron interactions.
Short-ranged interactions for quadratic dispersion in two dimensions are marginal\cite{sun,vafek,fan}
at the tree level while Coulomb 
interactions are relevant; this already points to the possibility of non-Fermi liquid behavior.
Unlike neutral graphene where electron-electron interactions are unscreened, Coulomb interactions
in bilayer graphene {\em are} screened due to the presence of a finite density of states, and dynamically
generate a momentum scale $\lambda = me^2/\varepsilon $ proportional to the inverse Thomas-Fermi
screening length $q_{TF}= 4 \lambda \log[4]$. Below this scale this system resembles one with 
effective short-ranged interactions. Based on the scaling form of the 
density-density response function we demonstrate within the Random Phase Approximation (RPA)
that there is {\em no} renormalization to the electron effective mass, while the imaginary part 
of the electron self-energy $\mathcal{I}m \Sigma \sim
\omega$ below the screening scale. As a result the quasiparticle has a logarithmically
vanishing spectral weight at low-energies, indicative of
non-Fermi liquid behavior. This is very different from a
2DEG where the phase space available for scattering is limited by
the energy thereby giving energy squared dependence for the quasiparticle decay rate, and the 
quasiparticle weight remains finite at the Fermi surface.
We argue this non-Fermi liquid behavior is a consequence of
the absence of a Fermi surface for neutral bilayer graphene and corresponding larger 
phase space available for scattering processes, which should be a robust result beyond RPA. The 
long-wavelength behavior of the electron spectral function
$A(k,\omega)=A(k, -\omega)$ is plotted in Fig.~\ref{spectralfunction}, which can be compared 
with ARPES measurements currently being employed to study the effects of interactions in
graphene systems~\cite{arpes_expt}. \newline

{\em Bilayer Graphene Effective Model} --- The low energy properties of Bernal stacked
bilayer graphene can be adequately described by quasiparticles with parabolic
dispersion~\cite{falko} exhibiting a Berry phase of 2$\pi$. Neglecting trigonal warping the
band hamiltonian for balanced bilayer graphene is:
\begin{equation}
\label{bandham}
{\hat {\cal H}} =  \sum_{\vec{k},\alpha} \frac{\vec{k}^2}{2 m} 
{\hat{\psi}}^{\alpha,\dagger}_{\vec{k}} [\tau^{z} \otimes (\vec{\sigma} \cdot
\hat{n}_{\vec{k}})]{\hat
\psi}^{\alpha}_{\vec{k}},
\end{equation}
where the Pauli matrix $\tau^{z}$ acts on the two-degenerate (K and K') valleys, $\vec{k}$ is
two-dimensional envelope function momentum measured from the two nodal points K and K',
$\sigma^{1}$ and $\sigma^{2} $ are  Pauli matrices that act on bilayer graphene's 
pseudospin (layer) degrees of freedom, and $\alpha = \uparrow,(\downarrow)$ accounts for the spin 
degrees of freedom. The chirality of bilayer graphene's chiral 2DEG is captured by the unit 
vector $\hat{n}_{\vec{k}} = (\cos 2 \varphi_{\vec{k}},\sin 2 \varphi_{\vec{k}})$ where  $\varphi_{\vec{k}}
= \tan^{-1}(k_{y}/k_{x})$. In Eq.~(\ref{bandham}) the field operator ${\hat
\psi}^{\alpha,\dagger}_{\vec{k}} = ({\hat \phi}^{\alpha,\dagger}_{K+\vec{k},t}
,{\hat \phi}^{\alpha,\dagger}_{K+\vec{k},b}, {\hat \phi}^{\alpha,\dagger}_{K'+\vec{k},b},
{\hat \phi}^{\alpha,\dagger}_{K'+\vec{k},t})$ is a four-component spinor where the low energy
sites\cite{falko} are the top (t) and bottom (b) layer sites without a near-neighbor in the
opposite layer. The effective mass is
determined by $m = \gamma_{1}/2v^2 \sim 0.054 m_{e}$ where $v$ is the single-layer Dirac 
velocity and $ \gamma_{1} \sim$ 0.4eV~\cite{graphite,falko} is the inter-layer hopping 
amplitude. \newline
\indent
The interaction contribution to the bilayer graphene's
hamiltonian is layer-dependent:
\begin{equation}
\label{eeham}
\mathcal{V} = \frac{1}{2L^2}
\sum_{\vec{q}} \sum_{\alpha,\alpha'} \big[ v_{+}(q) \hat{\rho}_{-q}^{\alpha}
\hat{\rho}_{q}^{\alpha'} + 2 v_{-}(q) \hat{\mathcal{S}}^{z}_{-q,\alpha}
\hat{\mathcal{S}}^{z}_{q,\alpha'}\big],
\end{equation}
where ${\hat \rho}_{q}^{\alpha} = \sum_{\vec{k}} \hat{\psi}^{\dagger \alpha}_{\vec{k}+\vec{q}}
\hat{\psi}^{\alpha}_{\vec{k}}$ is the total density per spin, $
\hat{\mathcal{S}}^{z}_{q,\alpha} = 1/2(\hat{\rho}^{\uparrow}_{q,\alpha} -
\hat{\rho}_{q,\alpha}^{\downarrow})$ is the z-component of the corresponding pseudospin density
operator in the K valley, with $ \hat{\mathcal{S}}^{z}_{q} \to - \hat{\mathcal{S}}^{z}_{q} $ in
the K' valley, and $v_{\pm}$ are the symmetric and antisymmetric combinations of the interaction
potentials in the same (different) layers $v_{s} = 2 \pi e^2/\varepsilon q  (v_{d} = 
v_{s}e^{-qd})$ with the layer separation $d = 0.334 nm$. \newline
\indent
{\em Response Functions} --- The
finite temperature non-interacting response functions can be written as
\begin{eqnarray}
\label{genresp}
&&\chi^{0}_{\alpha \alpha}(q,i\Omega_{n}) =\\ \nonumber &&- g \sum_{ss'} \int
\frac{d^{2}k}{(2 \pi)^{2}} \frac{n_{F}(\epsilon_{s}(\vec{k}))-
n_{F}(\epsilon_{s'}(\vec{k}+\vec{q}))}{i \Omega_{n} + \epsilon_{s}(\vec{k}) -
\epsilon_{s'}(\vec{k}+\vec{q})} \mathcal{F}^{ss'}_{\alpha \alpha}(\vec{k},\vec{k}+\vec{q}),
\end{eqnarray}
where $g=4$ accounts for the spin and valley degeneracy, $\alpha= \rho,
\mathcal{S}^{z}$ denotes the density and pseudospin density response functions, $s,s'= \pm$ is
the chiral band index, $\epsilon_{s}(\vec{k}) = sk^2/(2m) $, $n_{F}(\epsilon_{s}(\vec{k}))$ is
the Fermi-Dirac distribution and $\theta_{\vec{k},\vec{k}+\vec{q}}$ is the angle between the
wavevectors $\vec{k} $ and $\vec{k}+\vec{q}$.  The angular dependent matrix element
for the density-density response $\mathcal{F}^{ss'}_{\rho
\rho}(\vec{k},\vec{k}+\vec{q}) = 1/2(1+ss'\cos 2\theta_{\vec{k},\vec{k}+\vec{q}})$ is different
from the pseudospin density response $\mathcal{F}^{ss'}_{zz}(\vec{k},\vec{k}+\vec{q}) =
1/2(1-ss' \cos 2 \theta_{\vec{k},\vec{k}+\vec{q}})$ due to the presence of the pseudospin
operator $\hat{\mathcal{S}}^{z}_{q} \propto \tau^{z}$ in (\ref{eeham}). Physically these form
factors determine the relative weight of the interband ($ss'=-1$) and intraband ($ss'=+1$)
excitation contribution to the density-density and pseudospin density response functions. \newline
\indent
For undoped bilayer graphene where the Fermi energy lies at the neutrality point (i.e
$\epsilon_{F} =0$), the zero-temperature density-density response is completely determined by
interband excitations where the product $ss' =-1$. Our low-energy theory has a natural
high-energy momentum cutoff given by the bandwidth $k_{c}=\sqrt{2m\gamma_{1}}$, from
dimensional analysis of (\ref{genresp}) it is clear that the zero-temperature density-density
response function $\chi^{0}_{\rho \rho} = g m/(2\pi) \Phi_{\rho \rho}(q/k_{c},m i
\Omega_{n}/q^2)$, where $g m/(2\pi)$ is the constant density of states for parabolic dispersion,
and $\Phi_{\rho \rho}$ is a dimensionless scaling function. The density-density response function
is free of any divergences, and sending the bandwidth $k_{c} \to
\infty $ just gives $\chi^{0}_{\rho \rho} = g m/(2\pi) \Phi_{\rho \rho}(m i \Omega_{n}/q^2)$.
Analytically continuing $i \Omega_{n} \to \Omega + i \eta$ the respective real and imaginary
parts of scaling function $\Phi_{\rho \rho}$ are~\cite{castroneto} ($y=m\Omega/q^2$):
\begin{eqnarray}
\nonumber
\mathcal{R}e \Phi_{\rho \rho}(y) &=& \log[4] + \frac{1}{2y}
\log\big|\frac{1+2y}{1-2y} \big| - \frac{1}{4y} \log\big|\frac{1+4y}{1-4y} \big| \\ &+&
\log\big|\frac{1-4y^2}{1-16y^2} \big|, \\
\nonumber \mathcal{I}m \Phi_{\rho \rho}(y) &=& \pi
(1-\frac{1}{4y}) \Theta (y-\frac{1}{4}) - \pi (1-\frac{1}{2y}) \Theta (y-\frac{1}{2}),
\end{eqnarray}
valid for $\Omega \geq 0$. It is important to note that the above expression is a function of
a single scaling variable $y=m\Omega/q^2$, contrary to the case of any two-dimensional system 
with a Fermi wavevector $k_{F}$ where it is a function of two variables namely $m\Omega/k^{2}_{F}$ 
and $q/k_{F}$. This scaling behavior is a consequence of the absence of a Fermi surface for neutral 
bilayer graphene.  \newline
\indent
The imaginary part of the density-density response
function ($\mathcal{I}m \chi_{\rho \rho}(q,\Omega) = 0 $ for $\Omega > q^{2}/(4m)$), defines the
edge of the particle-hole continuum. For interband excitations this is just given by the minimum
and maximum values of $\Omega = \epsilon_{\vec{k}+\vec{q}} + \epsilon_{\vec{k}} $. The minimum
energy for a particle-hole pair is attained for $\theta_{\vec{k},\vec{q}}= \pi$, this can be seen
by completing the square and writing min($\Omega) = 1/m[(k-q/2)^2+q^2/4]$. The excitation
spectrum here $\Omega > q^{2}/(4m)$ is similar to that of neutral graphene $\Omega > v|q|$ with
the difference coming from switch to a parabolic dispersion. \newline
\indent
Within the Random Phase Approximation (RPA)~\cite{largegfootnote}:
\begin{equation}
v_{+}^{RPA}(q,\Omega) = \frac{v_{+}(q)}{1+v_{+}(q) \chi^{0}_{\rho \rho}(q,\Omega)},
\end{equation}
we recover the static screening ($\Omega \to 0$ limit of $v^{RPA}(q,\Omega)$)
form of Ref. \onlinecite{hwang}, with the Thomas-Fermi wavevector $q_{TF} = g (m e^2)/\varepsilon \log[4]$.
Due to the positive definite value of $\mathcal{R}e \chi^{0}_{\rho \rho}(q,\Omega)$, bilayer
graphene at the neutrality point excludes any plasmon excitations. The absence of a plasmon mode
is not unique for neutral bilayer graphene and is phenomenologically similar to the case of
neutral single-layer graphene. In neutral single-layer graphene the density-density response
function vanishes inside the particle-hole continuum thereby excluding plasmon excitations,
technically different from the case of bilayer graphene. Using the continuity equation we can
relate the optical conductivity to the density-density response function. The RPA optical
conductivity (accounting for spin and valley degeneracy):
\begin{equation}
\sigma(\Omega) = 4
\lim_{q \to 0} \frac{i e^2 \Omega}{q^2} \frac{\chi^{0}_{\rho \rho}(q ,\Omega)}{1+v_{q}
\chi^{0}_{\rho \rho}(q ,\Omega)} = \frac{e^2}{2 \hbar},
\end{equation}
(restoring $\hbar$) is purely real and frequency
independent similar to the case of single layer graphene. The optical conductivity 
is not influenced by interactions within RPA and retains its universal value of $e^2/2 \hbar$.   
\begin{figure}[t,l]
\begin{center}
\includegraphics[clip,width=3.375in,height=2.25in]{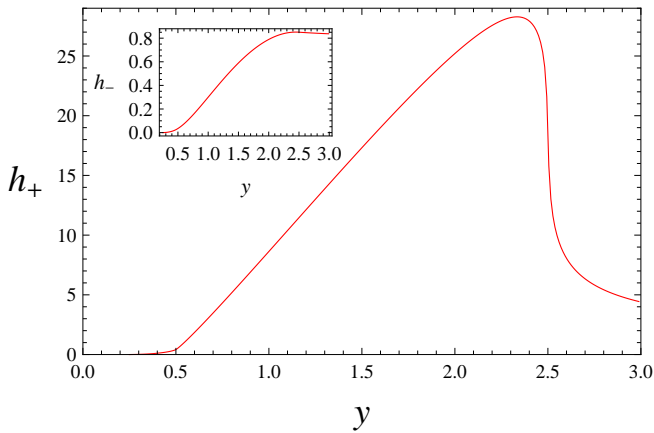}
\caption{(Color online) Scaling function $h_{+}(y) $ for the imaginary
part of the electron self-energy with Coulomb interactions
in the long-wavelength limit plotted as a function of $ y = m \omega/k^2$. 
The inset show the scaling function $h_{-}(y)$ also plotted as a function of $y = 
m \omega/k^2$. (see text for details).}
\label{scalingfunction}
\end{center}
\end{figure}

{\em Quasiparticle properties at $T=0$} ---
The essential feature of a Fermi liquid is encoded in the imaginary part of the quasiparticle
self-energy ($\mathcal{I}m \Sigma$) or inverse quasiparticle lifetime, which for a homogeneous 2DEG
gives $\mathcal{I}m \Sigma \sim (\epsilon - \epsilon_{F})^{2} \log|\epsilon - \epsilon_{F}|$. In the
case of neutral bilayer graphene the absence of a Fermi surface coupled with the effect of the
interband excitations invalidates such a description. In this section we calculate the imaginary
part of the quasiparticle self-energy for neutral bilayer graphene and show that $\mathcal{I}m
\Sigma \propto \epsilon_{\vec{k}}$ for $\epsilon_{\vec{k}} \to 0$ indicating non-Fermi liquid
behavior. \\
\indent
First let us examine Fermi's golden rule expression for short-ranged ($q$-independent)
interaction $u_{eff}$.
This is a rather crude approximation but as we argue below still gives the essential features of
the predicted non-Fermi liquid behavior. Based on Fermi's golden rule the inverse quasiparticle
lifetime can be expressed as:
\begin{equation}
\label{selfeneshort}
\frac{1}{\tau_{+,\vec{k}}} = \frac{g m u_{eff}^{2}}{2 \pi}\int \frac{d^{2} k'}{(2
\pi)^2} Im \Phi \left(\frac{m(\epsilon_{k'}-\epsilon_{k})}{|\vec{k}-\vec{k'}|^2}\right) 
\cos^{2} \theta_{\vec{k},\vec{k'}}.
\end{equation}
The $\epsilon_{k}$-dependence can be readily obtained by dimensional analysis which gives
$1/\tau_{+,\vec{k}} \propto \epsilon_{k}$; an exact
calculation yields $1/\tau_{+,\vec{k}} = 0.1076 g m u_{eff}^{2} \epsilon_{k}$.
The linear energy behavior only depends on the fact that $\Phi$ is a 
function of a single scaling variable $y = m \Omega/q^2$, which is a consequence of the 
scale invariance of neutral bilayer graphene, and is independent of the detailed behavior of 
$\Phi$. Bare Coulomb interactions would give a different result, however screening 
dynamically generates a new scale $q_{TF}$, below which interactions effectively behave as short 
ranged. Based on the effects of screening and independence of the inverse lifetime on the 
function $\Phi$ we anticipate non-Fermi liquid behavior in the long-wavelength limit. \\
\indent
The retarded quasiparticle self-energy within RPA can be written as:
\begin{equation}
\Sigma^{ret}_{s}(\vec{k},\omega) = \Sigma^{res}_{s} (\vec{k},\omega) +
\Sigma^{line}_{s}(\vec{k},\omega),
\end{equation}
following the the line and residue decomposition of Quinn and Ferrell~\cite{quinnandferrell}. The
quasiparticle self-energy within RPA remains diagonal in the particle-hole basis. It can be shown
that the line contribution is purely real and does not contribute to the imaginary part of
$\mathcal{I}m \Sigma^{ret}_{+}$ which for $(\omega >0)$:
\begin{eqnarray}
\label{imself}
\nonumber
\mathcal{I}m \Sigma^{ret}_{+}(k, \omega) &=& \sum_{s'= \pm} \int \frac{d^{2}q}{(2
\pi)^{2}} v_{+}(|\vec{k}-\vec{q}|)
\bigg( \frac{ 1 + s' \cos(2 \theta_{k,q})}{2} \bigg) \\
& & Im \big[\frac{1}{\varepsilon(|\vec{k}-\vec{q}|, \omega -
\epsilon(q))} \big] [\Theta( \omega -\epsilon(\vec{q}))],
\end{eqnarray}
where $\varepsilon(q,\Omega) = 1 + v_{+}(q) \chi_{\rho \rho}(q,\Omega)$.
In the above expression we have neglected the contribution of $v_{-}$ as it is 
logarithmically suppressed once screening effects are accounted for within RPA. 
Dimensional analysis of (\ref{imself}) implies that $\mathcal{I}m  \Sigma_{+} (\vec{k}, \omega) 
= |k|^2 f(m \omega/|k|^2,k/q_{TF})$,
where $f$ is a two variable function. Interactions introduce
an inverse length scale $\lambda = me^2/\varepsilon $ which turns out to be the same order of 
magnitude as the bandwidth cutoff $k_{c}$. In the long-wavelength limit we find that:
\begin{equation}
\label{selfenecoul}
\mathcal{I}m \Sigma^{ret}_{+} (\vec{k} \to 0,\omega \to 0) = \frac{|k|^{2}}{g 2\pi  m} h_{+}
(\frac{m\omega}{k^2}) \Theta (4m\omega -k^2).
\end{equation}
The scaling function $h_{+}(m\omega/k^{2}) $ was numerically attained and is plotted in
Fig.~\ref{scalingfunction}.
The theta function in (\ref{selfenecoul}) comes from the particle-hole continuum and is independent
of the nature of interactions. Using the symmetry relations $\mathcal{I}m \Sigma^{ret}_{s}
(\vec{k},- \omega ) = \mathcal{I}m \Sigma^{ret}_{\bar{s}}  (\vec{k}, \omega )$ where $\bar{s} = -s$
it is clear to see that  $\mathcal{I}m \Sigma^{ret}_{+}  (\vec{k}, -\omega )$ gives a similar 
expressions as~(\ref{selfenecoul}) with a different scaling function $h_{-}(m\omega/k^2)$ plotted in 
the inset of Fig.~\ref{scalingfunction}. The residue contribution to the real part of the retarded 
electron
self-energy $\mathcal{R}e \Sigma^{ret}_{+}$ yields a similar expression, however the contribution
due to $\Sigma^{line}_{s}$ is more singular thereby dominating in the long-wavelength limit. We
find that ($\omega >0 $):
\begin{equation}
\mathcal{R}e \Sigma^{ret}_{s} (\vec{k},\omega) = s\frac{2k^{2}}{g \pi^2 m}
(\log \big[ \frac{\lambda}{k} \big])^2 -  \frac{4 \omega}{g\pi^2} (\log \big[
\frac{\lambda}{\sqrt{m \omega}}])^2 + ...,
\end{equation}
where "..." represent the subleading terms. The expression for $\omega <0$ can be attained by
exploiting the symmetry relations $ \mathcal{R}e \Sigma^{ret}_{s} (\vec{k},-\omega) = -\mathcal{R}e
\Sigma^{ret}_{\bar{s}} (\vec{k},\omega)$.  The quasiparticle spectral weight:
\begin{equation}
\lim_{k \to 0}Z_{+} = \frac{1}{1- \partial_{\omega} \mathcal{R}e \Sigma_{+}^{ret}(k,\omega)} \sim
\frac{g \pi^2}{4} \bigg(\log\big[ \frac{\sqrt{2m^{\star}}}{m} \frac{\lambda}{k} \big]
\bigg)^{-2},
\end{equation}
vanishes logarithmically, whereas the effective mass
$m^{\star}$
\begin{equation}
\frac{m_{+}}{m^{\star}_{+}} = \frac{1+m_{0} \partial_{k^2} \mathcal{R}e
\Sigma^{ret}_{+}(k,\omega)}{1- \partial_{\omega} \mathcal{R} e \Sigma_{+}^{ret}(k,\omega)} \to 1,
\end{equation}
remains finite and is not renormalized by interactions. The long-wavelength behavior of the spectral
function $A(\vec{k},\omega) = A_{+} (\vec{k},\omega) + A_{-}(\vec{k},\omega)$ plotted in
Fig.~\ref{spectralfunction} with:
\begin{equation}
A_{s}(\vec{k},\omega) = \frac{1}{\pi} \frac{\mathcal{I}m \Sigma^{ret}_{s}
(\vec{k},\omega)}{|\omega - \epsilon_{k} - \mathcal{R}e \Sigma^{ret}_{s}
(\vec{k},\omega)|^2 +|\mathcal{I}m \Sigma^{ret}_{s} (\vec{k},\omega) |^2},
\end{equation}
was calculated from the leading order behavior of the electron self-energy and neglecting 
regular contributions. Symmetry relations for the electron self-energy dictate that $A(k,\omega) = 
A(k,-\omega)$.

\indent
The linear dependence of the $Im \Sigma$ on the quasiparticle energy predicted above is
different form the case of neutral single layer graphene. In neutral single layer graphene due
to the lack of screening associated with the Dirac point the Fermi velocity develops a
logarithmic enhancement~\cite{guineaprbrapid}. For neutral graphene this logarithimic velocity
enhancement implies that $Im \Sigma(\omega) \sim \omega/(\log \omega)^2$ which is smaller than
$Im \Sigma(\omega) \sim \omega $. In contrast to single layer graphene
interactions in bilayer graphene are screened, as we have shown and has been pointed out in the
literature~\cite{borghi,eefalko} with Thomas-Fermi screening, that the
quasiparticle dispersion in bilayer remains quadratic. Most
of our analysis of quasiparticle properties in neutral bilayer graphene has relied on the fact
that $ \chi^{0}_{\rho \rho} (q,\Omega) \propto \omega^{0}$ (i.e. it has a scaling behavior
after sending the bandwidth $k_{c}
\to \infty$). In the next section we identify this regime of non-Fermi liquid behavior at finite
temperature.\\

{\em Regime of Non-Fermi liquid behavior} ---
Weak interlayer hopping leads to trigonal warping of the band structure in bilayer
graphene~\cite{falko}. The temperature associated with this effect can be estimated by
calculating the energy scale at which trigonal warping effects compete with the quadratic
dispersion kept within our model. Using the bare parameters of graphite $\gamma_{3} \sim
0.1 \gamma_{0}$~\cite{graphite} we can estimate that the temperature below which the trigonal
warping effect becomes relevant is $T_{1} \sim 40K$. Below this scale the electron dispersion
crosses over from quadratic to linear, and the system behaves like single-layer graphene,
and our results no longer apply. Recently there has been discussions of interaction-driven
spontaneous symmetry breaking in neutral bilayer graphene in the absence of trigonal warping,
due to the marginal relevance of weak short-range repulsive interactions\cite{vafek,fan}.  Since 
the interaction is only marginally relevant, the transition temperature $T_c$ into the possible  
broken symmetry phases are exponentially small. The non-Fermi liquid behavior discussed here 
thus applies to temperatures above the higher of $T_c$ and $T_1$.
Our analysis can be extended to finite temperatures, again dimensional analysis dictates that
the polarization function have the scaling form $\Phi(m\Omega/k_{\beta}T,m \Omega/q^{2})$.
From a simple scaling analysis of the electron self-energy we conjecture that for
temperatures $T > max(T_{1},T_{c})$:
\begin{equation}
\mathcal{I}m \Sigma \sim \bigg{\{}
\begin{array}{c} \omega \qquad  \omega >k_B T \\
k_B T  \qquad  k_B T >  \omega
\end{array}
\end{equation}
indicative of non-Fermi liquid behavior. In this paper we have analyzed the regime $k<q_{TF}$, in 
the opposite regime $k>q_{TF}$ one can show that the non-Fermi behavior becomes even 
more pronounced as screening effect is less significant.

We acknowledge discussions with A. H MacDonald, V. Falko, S. Das Sarma, F.
Guinea, E. Hwang, O. Vafek and especially Fan Zhang for his comments on the broken symmetry phases.
This work was supported in part by NSF grant DMR-0704133 (KY)
and the State of Florida (YB).


\begin{thebibliography}{99}
\bibitem{largegfootnote}
Formally RPA can be justified with a Large-g expansion (commonly known as Large-N), becoming 
asymptotically exact for $g \to \infty$.
\end{thebibliography}
\end{document}